\documentclass[fleqn,10pt]{wlscirep}
\usepackage[utf8]{inputenc}
\usepackage[T1]{fontenc}
\usepackage{multirow}
\usepackage{threeparttable}
\usepackage{tikz}
\usetikzlibrary{arrows.meta, positioning, calc}

\title{Knowledge-guided Transfer Prediction In Underrepresented Populations: A GRU-D-Static Framework For Maternal And Neonatal Outcomes}

\author[1]{Yipeng Wei}
\author[2]{Zahra Hoodbhoy}
\author[3]{Emily R. Smith}
\author[1]{Fang Jin}
\author[2]{Muhammad Imran Nisar}
\author[2]{Muhammad Farrukh Qazi}
\author[3]{Christopher Mores}
\author[4]{Victor Akelo}
\author[4]{Caleb Sagam}
\author[4]{Florence Aweyo}
\author[5]{Charlotte Tawiah}
\author[5]{Veronica Agyemang}
\author[5]{Kwaku Poku Asante}
\author[5]{Sam Newton}
\author[6]{Santosh Joseph Benjamin}
\author[6]{Anne George Cherian}
\author[6]{Devakumar  Devadhas}
\author[6]{James A}
\author[7]{Margaret P. Kasaro}
\author[7]{Augustine Tunga}
\author[8]{Sarmila Mazumder}
\author[8]{Neeraj Sharma}
\author[9]{Wilbroad Mutale}
\author[10]{Mae Bridget Spelke}
\author[11,*]{Qing Pan}

\affil[1]{George Washington University, Department of Statistics, Washington D.C., United States}
\affil[2]{The Aga Khan University, Department of Paediatrics and Child Health, Karachi, Pakistan}
\affil[3]{George Washington University, Department of Global Health, Washington D.C., United States}
\affil[4]{Kenya Medical Research Institute, Centre for Global Health Research, Kisumu, Kenya}
\affil[5]{Kintampo Health Research Center, Research \& Development Division, Kintampo, Ghana}
\affil[6]{Christian Medical College \& Hospital, Vellore, India}
\affil[7]{University of North Carolina Global Projects Zambia, Lusaka, Zambia}
\affil[8]{Implementation Science Domain, Society for Applied Studies, New Delhi, India}
\affil[9]{University of Zambia, Department of Health Policy and Management, Lusaka, Zambia}
\affil[10]{University of North Carolina at Chapel Hill, Chapel Hill, United States}
\affil[11]{George Washington University, Department of Biostatistics and Bioinformatics, Washington D.C., United States}

\affil[*]{Corresponding Author: qpan@gwu.edu}

\begin{abstract}
Integrating summary-level scientific knowledge (for example, published external literature) into neural network models provides a practical strategy for transferring prediction models trained on adequately sampled source cohorts to underrepresented target populations, where individual-level data in the target domain are often limited or unavailable. In this study, we propose transfer prediction strategies incorporating external summary-level scientific knowledge and illustrate its application on the PRISMA Maternal and Neonatal Health Study, training a neural network model on the source data to predict adverse outcomes in the target cohorts. Besides, we also extend the existing GRU-D framework by incorporating static feature embeddings and attention weights to jointly leverage temporal and static information for improved prediction. Our approach employs soft labels derived from summary-level statistics (such as coefficients in logistic regressions) describing the target population to fine-tune GRU-D-Static models that are initially trained on the source populations which differ from the target population. We evaluate six maternal and neonatal outcomes, including stillbirth, preterm birth, low birth weight, small vulnerable newborn, neonatal death, and maternal near miss. Across all tested scenarios, fine-tuning using soft labels from just basic covariates substantially improved predictive performance compared with deep learning models trained on the source sample. Furthermore, the performance slightly improves more when additional covariates were incorporated into the logistic regression model or when partial input features from the target population were available for fine-tuning. These findings demonstrate that integrating existing scientific knowledge in the literature through transfer prediction of source neural network models can enhance prediction performance in underrepresented target populations, reducing reliance on large-scale data collection and supporting risk prediction in global health.
\end{abstract}
\begin{document}

\flushbottom
\maketitle
%
%
\thispagestyle{empty}

\section*{Introduction}
Transfer prediction aims to make inference or predictions for a target population by adapting models trained on related source populations, which is especially valuable in global healthcare and clinical studies, as data availability and population characteristics often differ substantially across regions \cite{zhuang2020comprehensive,iman2023review}. Large-scale global analyses have demonstrated substantial regional inequalities in maternal and neonatal health outcomes across LMICs, reflecting differences in both risk factor distributions and healthcare systems \cite{peng2024global}. In practice, the distribution of risk factors and the prevalence of outcomes can vary between regions. Furthermore, the relationships between them may also vary, such that directly applying a source-trained model can lead to poor prediction performance in target populations. For example, consider a neonatal risk prediction model developed in South Asian low- and middle-income countries (LMICs), such as India or Pakistan, using routinely collected medical data, including maternal demographics, clinical history, and basic laboratory measurements. When this model is applied to African settings such as Kenya, Zambia, or Ghana, its performance may deteriorate due to differences in the prevalence of infectious diseases (e.g., malaria, HIV), maternal nutritional status, and access to diagnostic resources. Prediction models often fail in underrepresented populations. Transfer prediction can help, but usually requires more target-domain data than are realistically available. In many real settings, only summary-level target information may be available. Existing longitudinal prediction frameworks do not adequately address this setting, especially when data are irregular and include static and temporal predictors. Therefore, this paper proposes two major innovations: an extension of GRU-D model to emphasize static features and a transfer-learning strategy that uses summary-level target information via soft labels.

Previous studies have shown that weighting methods can help mitigate such discrepancies \cite{ling2023overview}. Recent work in healthcare has addressed this challenge through transfer prediction frameworks such as COMMUTE\cite{gu2023commute} and TransRF\cite{gu2022transfer}, which calibrate or ensemble models across heterogeneous populations to improve generalizability in underrepresented groups. Such methods require individual-level covariate information from the target population. However, in some cases, data from the target population may be limited or completely unavailable and only prior summary-level information from previous studies, such as known associations between outcomes and risk factors can be relied upon. For example, one may wish to adapt a type 2 diabetes (T2D) risk prediction model trained on U.S. participants from the Diabetes Prevention Program (DPP) Outcomes Study \cite{diabetes2002reduction} to a Chinese population by incorporating summary-level logistic regression coefficients reported in the Wuhan elderly cohort study\cite{liu2022predicting}. In this case, the Wuhan study publishes regression coefficients from a penalized logistic model, while individual-level de-identified DPP data are available through the National Institute of Diabetes and Digestive and Kidney Diseases (NIDDK) data repository.  Ma et al.\cite{ma2018risk} proposed the PRIME framework for electronic health record (EHR) risk prediction, which employs a posterior regularization technique to incorporate prior medical knowledge (e.g., disease–risk factor relationships) into deep learning models. Recent reviews emphasize the importance of developing broader methodological approaches to support generalisability and transportability. For example, Ploddi et al. provide a comprehensive overview of both data-driven and knowledge-driven strategies that can strengthen model performance across populations \cite{ploddi2024scoping}. Manke-Reimers et al. review applied studies of transportability methods and illustrate how these approaches have been implemented to improve external validity in diverse settings \cite{manke2025and}. Additionally, under the constraints of limited target data, especially when actual labels are absent, models can still be adapted by leveraging prior knowledge through soft labels which encode class probabilities rather than binary outcomes. These soft labels provide additional information that improves predictions in target populations \cite{yao2019heterogeneous}. This approach, known as knowledge distillation, has been shown to be effective for domain adaptation \cite{hinton2015distilling,willard2022integrating}.

The Pregnancy Risk, Infant Surveillance, and Measurement Alliance (PRISMA) Maternal and Newborn Health Study is a population-based, longitudinal observational study designed to collect standardized data for estimating maternal and neonatal risks and to develop innovative strategies to improve pregnancy outcomes for mothers and their newborns in low- and middle-income countries\cite{peng2024global}, including Kenya, Zambia, Ghana, Pakistan, and India. In this study, we divide the PRISMA data into African and Asian regions. Given the regional heterogeneity across Asia and Africa, our study explores transfer prediction approach using soft labels to fine-tune deep learning models from one region (source) to another (target) in the absence of training data from the target population. We focus on critical maternal (e.g., maternal near miss) and neonatal outcomes, including stillbirth, low birth weight (LBW), preterm birth (PTB), small vulnerable newborn (SVN), and neonatal mortality\cite{islam2022machine,mangold2021machine}. We first train the deep learning model on the source data and then fine-tune it using summary-level information of the relationships between the covariates and the disease outcome in the target region. That is, we generate soft labels using a logistic regression model fitted on the target population whose coefficients are available on public literatures. Transfer prediction is not strictly required for the PRISMA data, as detailed individual-level information on both covariates and outcomes is available for both the source and target populations. Nevertheless, we use the PRISMA data to illustrate the proposed transfer prediction approach, which is particularly useful in settings where only summary-level information is available for the target population alongside individual-level data from the source population. Notably, the Asian and African PRISMA cohorts were collected under a common protocol with harmonized variables and consistent quality standards. As a result, the distributional shift between the source and target populations is relatively limited, representing a favorable transfer prediction setting. The prediction performance of the fine-tuned deep learning model is subsequently evaluated on the target data in PRISMA to assess the effectiveness of the transfer prediction approach. 

We implement the GRU-D-Static deep learning framework, a variant of the GRU-D framework developed by our team in this study. Nevertheless, the proposed transfer prediction methodology is model-agnostic and broadly applicable to other deep learning frameworks. The original GRU-D framework is designed to handle multivariate clinical time series with missing values through a decay mechanism that adjusts input features and hidden states based on the time elapsed since the last observation \cite{che2018recurrent}. However, a key limitation of the GRU-D framework is that static features cannot be incorporated, as these features are typically measured once without any associated time intervals, in which case the decay mechanism cannot be applied. To overcome this limitation, we propose the GRU-D-Static framework which integrates static feature embeddings with attention weight that dynamically weights temporal information based on the static context to enhance predictive performance \cite{shickel2022multi}.

To the best of our knowledge, this is among the first approaches to leverage external scientific knowledge (e.g., logistic regression coefficients) as soft labels, rather than true outcome labels, for transfer prediction in data-limited target populations. The proposed method is well suited for low- and middle-income country (LMIC) contexts, where access to comprehensive individual-level data is often limited or unavailable.

\section*{Methods}
\subsection*{Data}
We select six maternal and neonatal outcomes to validate our transfer prediction approach: \begin{enumerate}
  \item Maternal near miss: Refers to women who experienced life-threatening complications but survived during pregnancy, childbirth, or within 42 days of the end of pregnancy based on WHO near miss clinical criteria.
  
  \item Stillbirth: Defined as the delivery of a fetus with no signs of life, indicated by the absence of breathing, heartbeat, umbilical cord pulsation, or voluntary muscle movement at or after 20-week gestation.
  
  \item Low birth weight (LBW): Defined as a birth weight less than 2500 grams.
  
  \item Preterm birth (PTB): Defined as live births occurring before 37 completed weeks of gestation.
  
  \item Small vulnerable newborn (SVN): Defined as a four-category composite based on gestational age and size-for-gestational-age: (1) Term + non-small for gestational age (non-SGA), (2) Term + small for gestational age (SGA), (3) Preterm + non-SGA, and (4) Preterm + SGA.
  
  \item Neonatal mortality: Defined as the death of a liveborn infant before 28 completed days of life.
\end{enumerate}

\begin{table}[htbp]
\centering
\scriptsize
\begin{tabular}{|p{2.5cm}|p{2cm}|p{1cm}|p{6cm}|}
\hline
\textbf{Category} & \textbf{Risk factor} & \textbf{Type} & \textbf{Lab Indicator, Data Source, Definition} \\
\hline
\multirow{2}{*}{\textbf{Sociodemographics}} 
& Demographics & Static & Years of school $\geq$ 10, Maternal age, Unimproved water source, Unimproved sanitation, Wealth index, Paid work \\
\cline{2-4}
& Air pollution & Static & Household smoking, Smoking, Chew tobacco, Chew betelnut, Use of unclean cooking fuel (kerosene, coal, charcoal, biomass, wood) \\
\hline
\multirow{6}{*}{\textbf{Clinical}} 
& Nutritional status & Static & Maternal BMI, Gestational weight gain (IOM guideline\cite{steinberg2011clinical}), Maternal MUAC \\
\cline{2-4}
& Medical history & Static & Preterm birth, C-section, Miscarriage $\geq$ 3, Stillbirth, Parity \\
\cline{2-4}
& \multirow{2}{*}{Comorbids} & Static & Chronic hypertension, Pregestational diabetes \\
\cline{3-4}
& & Static & Gestational hypertension (including preeclampsia), Gestational diabetes, Maternal depression \\
\cline{2-4}
& Delivery & Static & Place of delivery, GA at birth, Birthweight, Newborn gender \\
\cline{2-4}
& Postnatal & Static & Congenital anomalies, PSBI, Excessive TCB by NICE threshold \cite{amos2017jaundice}\\
\hline
\multirow{3}{*}{\textbf{Lab}} 
& \multirow{2}{*}{Infections} & Static & Hep B, Hep C \\
\cline{3-4}
& & Static & HIV/AIDS, Tuberculosis, Malaria infection, Other STIs \\
\cline{2-4}
& Biomarkers & Temporal & Hemoglobin \& Maternal anemia, Total iron binding capacity, Ferritin, Hepcidin \\
\hline
\multirow{3}{*}{\textbf{Ultrasound}} 
& \multirow{3}{*}{Ultrasound} & Static & Number of fetus \\
\cline{3-4}
& & Temporal & AFI index, Placental anomalies \\
\hline
\end{tabular}
\caption{Summary of risk factor categories and definitions}
\label{table:covariates}
\vspace{0.2cm}
\parbox{0.9\textwidth}{\scriptsize 
\textit{Abbreviations:} BMI, body mass index; MUAC, mid-upper arm circumference; GA, gestational age; IOM, Institute of Medicine; PSBI, possible serious bacterial infection; 
TCB, transcutaneous bilirubin; NICE, National Institute for Health and Care Excellence; 
STI, sexually transmitted infection; AFI, amniotic fluid index.
}
\end{table}

Potential risk factors incorporated in the models are categorized into temporal or static features (summarized in Table \ref{table:covariates}). Temporal features, are collected longitudinally at enrollment and at gestational weeks 20, 28, 32, and 36 \cite{wei2026_unpub}. Static features are collected only once. These risk factors include socio-demographic characteristics, clinical information, laboratory-related results and ultrasound findings. Basic covariates such as sociodemographic and clinical information are generally readily available in standard clinical settings. In contrast, access to laboratory testing and ultrasound examinations may be limited in low- and middle-income countries (LMICs) due to resource constraints. In particular, the detailed features included for each outcome are presented in Supplementary Table 1. For the preterm birth outcome, temporal features are restricted to those collected before 32 weeks of gestation to prevent information leakage.

\subsection*{Notations}
We denote the temporal features of length $T$ as $X_{temp}=(x_1, x_2,\dots, x_T)^T$ with corresponding temporal masking features $M_{temp}=(m_1, m_2,\dots, m_T)^T$. For each $t \in \{1, 2, \dots, T\}$, we define $x_t=(x_{t1}, x_{t2},\dots,x_{tU})^T$ and $m_t=(m_{t1}, m_{t2},\dots,m_{tU})^T$, where $U$ denotes the number of features at each time stamp. For each $d \in \{1, 2, \dots, U\}$, $m_{td}$ is defined as 
\begin{equation}
m_{td} = 
\begin{cases}
1, & \text{if } x_{td} \text{ is observed}\\
0, & \text{otherwise}
\end{cases}
\tag{1}
\end{equation}
The time stamps for temporal feature $X_{temp}$ are denoted by $S_{temp}=(s_1, s_2,\dots, s_T)^T$ with corresponding time intervals $\Delta_{temp}=(\delta_1, \delta_2,\dots, \delta_T)^T$, where $\delta_t = 0$ when $t=1$ and $\delta_t = s_t-s_{t-1}$ for $t>1$.\\
Similarly, the static features are denoted as $X_{static}=(x_{(s)1}, x_{(s)2},\dots,x_{(s)V})$ containing $V$ features measured once per subject. In this paper, we will predict the outcome $Y$ given a dataset of $N$ subjects, represented as $(X^{(n)},M^{(n)},S^{(n)})^N_{n=1}$, where $X^{(n)}=(X^{(n)}_{temp}, X^{(n)}_{static}), M^{(n)} = M^{(n)}_{temp} = (m^{(n)}_1, m^{(n)}_2,\dots, m^{(n)}_T)^T$ and $S^{(n)} = S^{(n)}_{temp} = (s^{(n)}_1,\dots,s^{(n)}_T)$ denotes the sequence of observation time points for subject $n$.
\subsection*{GRU-D-Static framework}
GRU-D is a specialized variant of the Gated Recurrent Unit (GRU) designed to model multivariate time-series data with missing values, where missingness may itself be informative \cite{che2018recurrent}.  
A key limitation of the original GRU-D framework is that it can only handle temporal data, as the decay mechanism requires the time interval vector $\delta_t$.  
To overcome this limitation, we introduce the GRU-D-Static framework, which combines static feature embeddings with an attention weight function that dynamically adjusts the contribution of temporal representations based on static context\cite{shickel2022multi}.

The component of the GRU-D-Static framework for handling temporal features follows the standard GRU-D formulation.  
At each time stamp $t$, the model takes the temporal input $x_t \in \mathbb{R}^U$, the masking vector $m_t \in \{0,1\}^U$, and the time interval vector $\delta_t \in \mathbb{R}^U$.  
A trainable input-decay is applied component-wise:
$$
\gamma_t = \exp\!\big(-\mathrm{ReLU}(W_\gamma \delta_t)\big) \in \mathbb{R}^U,
$$
and the decayed input used by the GRU cell is
$$
x^d_t = m_t \odot x_t + (1 - m_t) \odot \Big( \gamma_t \odot x'_t + (1 - \gamma_t) \odot \bar{x}_{t} \Big),
$$
where $\bar{x}_{t} = (\bar{x}_{t1}, \bar{x}_{t2},\dots,\bar{x}_{tU})^T$ is the empirical mean over all subjects, $x'_t = (x'_{t1}, x'_{t2},\dots,x'_{tU})^T $ is the last observed value carried forward, and $\odot$ denotes element-wise multiplication.  

The GRU gates are computed as:
$$
R_t = \sigma\!\left(W_R^{(x)} x^d_t + W_R^{(h)} H_{t-1} + W_R^{(m)} m_t + b_R \right),
$$
$$
Z_t = \sigma\!\left(W_Z^{(x)} x^d_t + W_Z^{(h)} H_{t-1} + W_Z^{(m)} m_t + b_Z \right),
$$
$$
\tilde{H}_t = \tanh\!\left(W^{(x)} x^d_t + W^{(h)} (R_t \odot H_{t-1}) + W^{(m)} m_t + b \right),
$$
$$
H_t = (1 - Z_t) \odot H_{t-1} + Z_t \odot \tilde{H}_t.
$$

Our GRU-D-Static framework incorporates static features $X_{\text{static}} \in \mathbb{R}^V$.  
These are first embedded via a fully connected layer with ReLU activation:
$$
\phi_{\text{static}} = \mathrm{ReLU}\!\big(W_{\text{static}} X_{\text{static}}\big),
$$
then generate element-wise attention weights over the final GRU hidden state:
$$
\alpha_{\text{static}} = \sigma\!\big(W_{\text{att}} \phi_{\text{static}}\big) \in \mathbb{R}^{\dim(H_t)}.
$$
The attention-weighted temporal representation is:
$$
H_t^{\mathrm{att}} = \alpha_{\text{static}} \odot H_t.
$$

Finally, $H_t^{\mathrm{att}}$ is concatenated with the static embedding, passed through dropout, and mapped to the output:
$$
H_{\text{final}} = \mathrm{concat}\!\big(H_t^{\mathrm{att}},\,\phi_{\text{static}}\big)
$$
$$
\hat{Y} = \sigma\!\big(W_{\text{out}} H_{\text{final}}\big)
$$

\begin{figure}[!ht]
    \centering
        \includegraphics[scale=0.7]{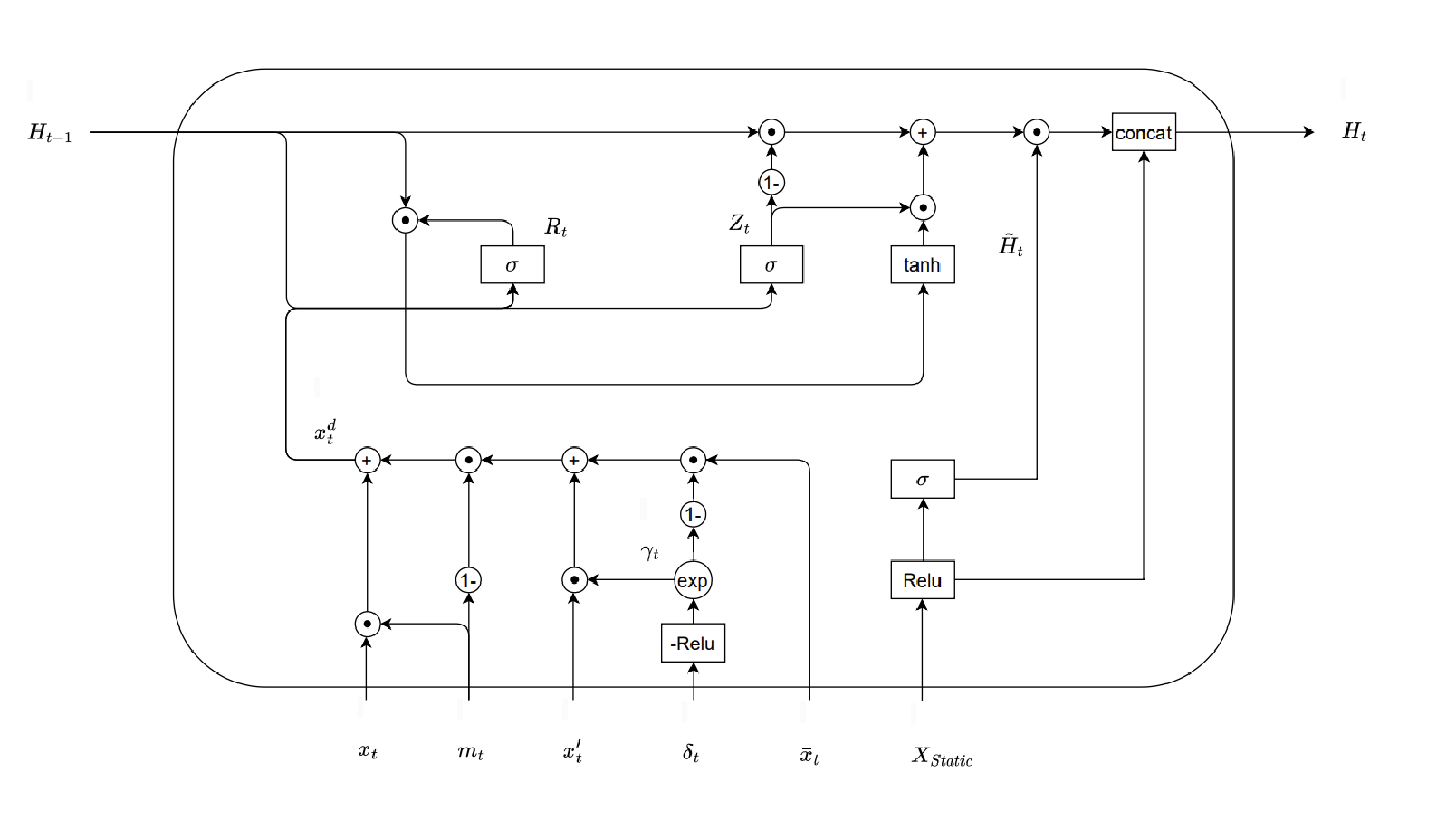}
        \caption{GRU-D-Static framework}
        \label{fig:GRU-D-Static}
\end{figure}

This architecture in Figure \ref{fig:GRU-D-Static} enables the static covariates to (i) reweight the temporal information via the attention vector $\alpha_{\text{static}}$ and (ii) contribute directly through $\phi_{\text{static}}$, while preserving GRU-D’s ability to model temporal dependencies and informative missingness in $X_{\text{temp}}$.

All features were preprocessed prior to modeling: binary features were coded as 1 (“Yes”) and –1 (“No”), and continuous features were normalized to a 0–1 range using min–max scaling after removing outliers \cite{patro2015normalization}.
For temporal features, missingness was explicitly modeled through masking features $m_t$ and time intervals $\delta_t$, which were incorporated via a trainable input-decay mechanism $\gamma_t$. Covariates collected after the occurrence of the outcome were deleted to prevent information leakage. For participants with features missing at all time points, as well as for static features, imputation was performed using region-specific empirical population means for continuous features to prevent data leakage across regions, and zeros for binary features. In this analysis, missingness across features was generally low, with a mean (SD) of 4.1\% (8.7\%), and the maximum missingness observed was 37.2\%.

\subsection*{Transfer prediction by integrating external scientific knowledge}
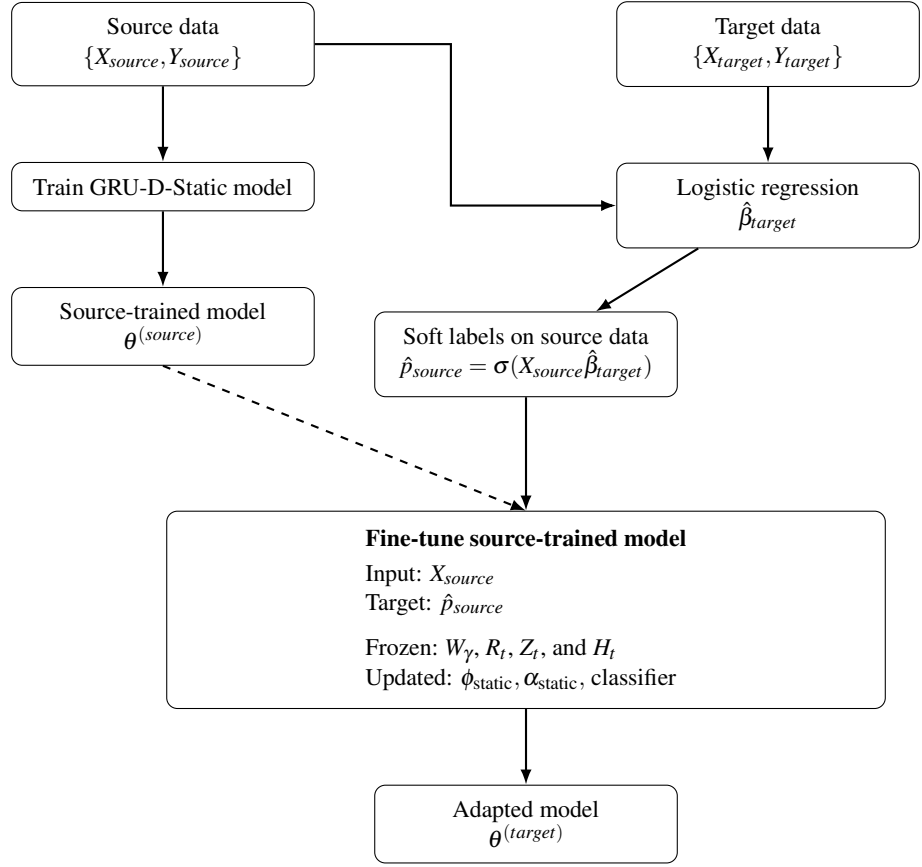
\begin{figure}[t]
\centering
\begin{tikzpicture}[
  font=\small,
  box/.style={draw, rounded corners, align=center, inner sep=6pt, minimum width=40mm},
  bigbox/.style={draw, rounded corners, align=left, inner sep=7pt, minimum width=95mm},
  arrow/.style={-{Latex[length=2mm]}, thick},
  dashedarrow/.style={-{Latex[length=2mm]}, thick, dashed},
  node distance=10mm and 18mm
]

\node[box] (srcdata) at (-4,2.3)
{Source data\\$\{X_{source}, Y_{source}\}$};

\node[box, below=of srcdata] (train)
{Train GRU-D-Static model};

\node[box, below=of train] (srcmodel)
{Source-trained model\\$\theta^{(source)}$};

\draw[arrow] (srcdata) -- (train);
\draw[arrow] (train) -- (srcmodel);

\node[box] (tgtdata) at (4,2.3)
{Target data\\$\{X_{target},  Y_{target}\}$};

\node[box, below=of tgtdata] (logreg)
{Logistic regression \\$\hat{\beta}_{target}$};

\draw[arrow] (tgtdata) -- (logreg);

\draw[arrow] (srcdata.east) -- ++(1.8,0) |- (logreg.west);

\node[box, below=15mm of $(train)!0.6!(logreg)$] (soft)
{Soft labels on source data\\
$\hat{p}_{source} = \sigma(X_{source} \hat{\beta}_{target})$};

\draw[arrow] (logreg) -- (soft);

\node[bigbox, below=15mm of soft] (finetune)
{\textbf{Fine-tune source-trained model}\\[2pt]
Input: $X_{source}$\\
Target: $\hat{p}_{source}$\\[6pt]
Frozen: $W_\gamma$, $R_t$, $Z_t$, and $H_t$\\
Updated: $\phi_{\text{static}}, \alpha_{\text{static}},$ classifier};

\node[box, below=of finetune] (adapted)
{Adapted model\\$\theta^{(target)}$};

\draw[arrow] (soft) -- (finetune);
\draw[arrow] (finetune) -- (adapted);

\draw[dashedarrow]
  (srcmodel.south) -- (finetune.north);
\end{tikzpicture}

\caption{Flowchart of transfer prediction framework for GRU-D-Static based on Scenario 1 and 2}
\label{fig:transfer_workflow}
\end{figure}

In real-world applications, source and target datasets often exhibit notable differences in risk factor distributions and outcome prevalence. When data from the target population are scarce or entirely absent, applying a model trained solely on the source data can lead to suboptimal predictive accuracy\cite{hosna2022transfer}. To address this challenge, we aim to integrate summary-level scientific knowledge from the target domain to enhance the performance of a source-trained model. The PRISMA study collected high-quality, large-scale data from both African and Asian sites. For illustration, we first treat the Asian data as the source population and the African data as the target population, under a hypothetical scenario in which individual-level data from the African sites are unavailable, while predictions are still required. We then repeat the analysis by reversing the roles of the source and target populations, treating the African sites as the source and the Asian sites as the target. To demonstrate the performance of the proposed transfer prediction approach, we evaluated model validity under the following hypothetical scenarios:
\begin{enumerate}
\item No individual-level target data are available, except for scientific knowledge regarding the association between the outcome and basic covariates.
\item No individual-level target data are available, except for scientific knowledge regarding the association between the outcome and all covariates.
\item Scientific knowledge is available, along with basic covariates information from the target data (labels are unavailable).
\end{enumerate}
To mimic the summary-level scientific knowledge available on external resources such as existing literatures, we fit a multivariate logistic regression model on the target data similar to publications on clinical and medical journals:
$$
Logit(Y_{target})=X_{target}\beta_{target}
$$ 
In Scenarios 1 and 2, we first use the estimated coefficients $\hat{\beta}_{target}$ to generate predicted probabilities $\hat{p}_{source}$ for the source data. These predicted probabilities serve as soft labels. Then we combine these soft labels $\hat{p}_{source}$, together with the covariates $X_{source}$ to fine-tune the source-trained GRU-D-Static framework. The workflow for Scenarios 1 and 2 is illustrated in Figure \ref{fig:transfer_workflow}. In Scenario 3, since partial target data is available, the estimated coefficients $\hat{\beta}_{target}$ are applied to the partial target covariates $X_{target}$ to produce soft labels $\hat{p}_{target}$. These soft labels along with the corresponding covariates $X_{target}$ are used to fine-tune the source-trained model. The prediction performance is evaluated on the target dataset.

In the fine-tuning process, the temporal parameters captured from the source data which includes the parameters of the decay mechanism $W_\gamma$ and the GRU gate functions $R_t$, $Z_t$, and $H_t$ are frozen. Instead of updating only the final classification layer, we also fine-tune the static embedding layer $\phi_{\text{static}}$ and the attention weight module $\alpha_{\text{static}}$. This design allows the model to adjust the influence of static features on temporal representations while refining their embeddings to reflect characteristics specific to the target population using soft labels derived from logistic regression models fitted on the target data, improving predictive performance on the target population\cite{zhuang2020comprehensive}. Our fine-tuning design follows other common transfer prediction methodology, in which general feature representations are fixed and the more specialized layers are adjusted to capture target-specific patterns \cite{yosinski2014transferable}.

\section*{Results}
In real-world applications, the most common situation corresponds to Scenario 1, where no target data are directly available except for existing summary-level scientific knowledge about associations between the outcome and basic covariates which are readily available. In the DPP-to-Wuhan example, the source and target cohorts come from different studies with distinct protocols, measurement approaches, and population characteristics. In contrast, the source and target cohorts in the PRISMA study share the same protocol, identical data collection instruments, consistent variable definitions, and comparable quality standards. Consequently, the transfer learning problem addressed here is substantially more straightforward than other real-world use case. Within the target data, 75\% are used to fit a logistic regression model and the remaining 25\% are reserved for performance evaluation. The logistic regression model includes basic covariates, such as socio-demographic characteristics and clinical information. The results without fine-tuning are summarized in the ``Baseline'' column of Table~\ref{table:source}, and the results with fine-tuning are shown in the ``Basic'' column. When the GRU-D-Static model is fine-tuned using the source covariates $X_{source}$ and soft labels $\hat{p}_{source}$ generated from the target logistic regression model, its predictive performance improves substantially. For instance, when the Asian data are treated as the source population and the African data as the target population, the AUC increases from 0.738 to 0.879 for stillbirth, from 0.574 to 0.602 for maternal near miss, and from 0.585 to 0.665 for neonatal death. Conversely, when the African data are treated as the source population and the Asian data as the target population, the AUC increases from 0.946 to 0.983 for stillbirth, from 0.592 to 0.668 for preterm birth, and from 0.693 to 0.868 for neonatal death.

To assess robustness of the proposed transfer prediction method, we additionally introduce noise to the external scientific knowledge by randomly flipping 10\% of outcomes in the target logistic regression model fitting to reflect the fact that summary-level associations in the literature may not match our target population 100\%. For example, publications often reflect data five to ten years ago while researchers want to make predictions for the current target population. As shown in the ``Noise'' column of Table~\ref{table:source}, although there is a small drop of AUC from the "Basic" column without noise, the fine-tuned model still outperforms the baseline for all outcomes, indicating that the proposed transfer prediction approach remains robust to moderate label perturbations.

To further assess the validity of our transfer prediction approach, we also examine Scenario 2, where no target data are directly available, except for summary-level scientific knowledge regarding the association between the outcome and all covariates. Similarly, within the target data, 75\% are used to fit a logistic regression model and the remaining 25\% are reserved for performance evaluation. The results are presented in the ``All'' column of Table \ref{table:source}. The fine-tuned GRU-D-Static framework again outperforms the source-trained GRU-D-Static framework. Compared with Scenario 1, fine-tuning with all covariates yields slightly better predictive performance over fine-tuning with only basic covariates. 

For comparison, we also report the AUC results of the GRU-D-Static model trained directly on the target data. Specifically, 75\% of the target data are used for model training and the remaining 25\% for performance evaluation. The results are presented in the ``Target'' column of Table~\ref{table:source}, which can be viewed as the upper-bound of the prediction performance for transfer prediction models. Except for the stillbirth outcome when African data are treated as target populations, the ROC-AUC values obtained from the model trained on the target data are not substantially different from those achieved under the fine-tuned Scenario 1 and Scenario 2, suggesting that the proposed fine-tuning strategy can approximate the performance of a model trained directly on the target domain.

In addition, we report the ROC-AUC results of logistic regression models fitted directly on the target data using the same set of basic covariates. The target dataset are again split into 75\% for fitting logistic regression and 25\% for testing, and the results are summarized in the ``LR'' column of Table~\ref{table:source}. The fine-tuned GRU-D-Static model achieved higher ROC-AUC values than logistic regression, indicating that our proposed transfer prediction approach provides better predictive performance than a conventional logistic regression method that directly applies summary-level scientific knowledge.

In Scenario 3, we evaluate transfer prediction when partial covariate information from the target data is available. Again, within the target data, 75\% are used to fit a logistic regression model and the rest 25\% are reserved for performance evaluation. A logistic regression model is fitted using the same basic covariates as in Scenario 1. Unlike Scenario 1, instead of using $X_{source}$ and $\hat{p}_{source}$, we fine-tune the GRU-D-Static framework with $X_{target}$ and the soft labels $\hat{p}_{target}$. We assess performance when 5\%, 10\%, 20\%, 50\%, and 100\% of the target data (used for fitting the logistic regression model) are employed for fine-tuning. As shown in Table \ref{table:target}, performance improves as more target data are incorporated. Notably, fine-tuning using only 10\% of the target data achieving performance comparable to that obtained with the full source dataset.

\begin{table}[htbp]
\centering
\footnotesize
\begin{threeparttable}
\caption{ROC-AUC on target data under Scenarios 1 and 2}
\label{table:source}
\begin{tabular}{l c c c c c c}
\toprule
\multicolumn{7}{l}{\textbf{Asia (Source) \& Africa (Target)}} \\
\midrule
\textbf{Outcome} & \textbf{Baseline} & \textbf{Basic} & \textbf{Noise} & \textbf{All} & \textbf{Target} & \textbf{LR} \\
\midrule
Maternal near miss (8.6\%)    & 0.574 & 0.602 & 0.595 & 0.608 & 0.616 & 0.584  \\
Stillbirth (2.7\%)           & 0.738 & 0.879 & 0.829 & 0.881 & 0.961 & 0.831  \\
Low birthweight (14.7\%)     & 0.830 & 0.842 & 0.831 & 0.848 & 0.847 & 0.832  \\
Small vulnerable newborn (25.2\%) & 0.641 & 0.662 & 0.659 & 0.668 & 0.684 & 0.637 \\
Preterm birth (8.2\%)        & 0.682 & 0.695 & 0.692 & 0.691 & 0.702 & 0.645 \\
Neonatal death (0.9\%)       & 0.585 & 0.665 & 0.669 & 0.672 & 0.666 & 0.619 \\
\midrule
\multicolumn{7}{l}{\textbf{Africa (Source) \& Asia (Target)}} \\
\midrule
\textbf{Outcome} & \textbf{Baseline} & \textbf{Basic} & \textbf{Noise} & \textbf{All} & \textbf{Target} & \textbf{LR} \\
\midrule
Maternal near miss (5.0\%)    & 0.637 & 0.657 & 0.654 & 0.661 & 0.673 & 0.634  \\
Stillbirth (1.3\%)           & 0.946 & 0.983 & 0.966 & 0.986 & 0.980 & 0.979  \\
Low birthweight (24.4\%)     & 0.808 & 0.838 & 0.831 & 0.835 & 0.835 & 0.834  \\
Small vulnerable newborn (34.4\%) & 0.641 & 0.644 & 0.646 & 0.651 & 0.662 & 0.623 \\
Preterm birth (16.7\%)       & 0.592 & 0.668 & 0.658 & 0.667 & 0.681 & 0.663 \\
Neonatal death (1.9\%)       & 0.693 & 0.868 & 0.857 & 0.864 & 0.878 & 0.858 \\
\bottomrule
\end{tabular}
\begin{tablenotes}
\scriptsize
\item \textit{Note:} Percentages in parentheses indicate the prevalence of outcomes in the target dataset.
\end{tablenotes}
\end{threeparttable}
\end{table}

\begin{table}[htbp]
\centering
\footnotesize
\begin{threeparttable}
\caption{ROC-AUC on target data under Scenario 3}
\label{table:target}
\begin{tabular}{l c c c c c c}
\toprule
\multicolumn{7}{l}{\textbf{Asia (Source) \& Africa (Target)}} \\
\midrule
\textbf{Outcome} & \textbf{Baseline} & \textbf{5\%} & \textbf{10\%} & \textbf{20\%} & \textbf{50\%} & \textbf{100\%} \\
\midrule
Maternal near miss (8.6\%)    & 0.574 & 0.588 & 0.587 & 0.597 & 0.593 & 0.589 \\
Stillbirth (2.7\%)           & 0.738 & 0.849 & 0.895 & 0.913 & 0.921 & 0.917 \\
Low birthweight (14.7\%)     & 0.830 & 0.842 & 0.835 & 0.845 & 0.840 & 0.845 \\
Small vulnerable newborn (25.2\%) & 0.641 & 0.655 & 0.655 & 0.652 & 0.663 & 0.657 \\
Preterm birth (8.2\%)        & 0.682 & 0.685 & 0.689 & 0.692 & 0.695 & 0.691 \\
Neonatal death (0.9\%)       & 0.585 & 0.648 & 0.662 & 0.660 & 0.651 & 0.656 \\
\midrule
\multicolumn{7}{l}{\textbf{Africa (Source) \& Asia (Target)}} \\
\midrule
\textbf{Outcome} & \textbf{Baseline} & \textbf{5\%} & \textbf{10\%} & \textbf{20\%} & \textbf{50\%} & \textbf{100\%} \\
\midrule
Maternal near miss (5.0\%)    & 0.637 & 0.644 & 0.659 & 0.653 & 0.657 & 0.655 \\
Stillbirth (1.3\%)           & 0.946 & 0.969 & 0.974 & 0.983 & 0.980 & 0.982 \\
Low birthweight (24.4\%)     & 0.808 & 0.827 & 0.832 & 0.836 & 0.834 & 0.837 \\
Small vulnerable newborn (34.4\%) & 0.641 & 0.639 & 0.649 & 0.643 & 0.651 & 0.654 \\
Preterm birth (16.7\%)       & 0.592 & 0.663 & 0.665 & 0.664 & 0.665 & 0.667 \\
Neonatal death (1.9\%)       & 0.693 & 0.807 & 0.837 & 0.862 & 0.863 & 0.872 \\
\bottomrule
\end{tabular}
\begin{tablenotes}
\scriptsize
\item \textit{Note:} Percentages in parentheses indicate the prevalence of outcomes in the target dataset.
\end{tablenotes}
\end{threeparttable}
\end{table}

\section*{Discussion}

In this study, we extend the existing GRU-D framework, which was originally developed to handle longitudinal data with missing values. A key limitation of the existing GRU-D framework is its inability to incorporate static features, as the decay mechanism relies on time intervals. To address this, we integrate static feature embeddings with attention weight, allowing static features to modulate temporal representations, and thereby improving predictive performance.

We further explore transfer prediction across heterogeneous populations by incorporating summary-level scientific knowledge from the target domain. Specifically, we fine-tune the source-trained model using soft labels generated from logistic regression models fitted on the target data. Although the GRU-D-Static framework is used in this study, the proposed transfer prediction strategy is model-agnostic and can be applied to other deep learning frameworks. This approach enables effective model adaptation even when individual-level outcome labels in the target population are unavailable. In Scenario 1, our results show that fine-tuning with both source covariates and soft labels yields substantial improvements in predictive accuracy, even when 10\% of outcomes are randomly flipped to introduce noise. In Scenario 2, slightly better performance when all covariates, rather than only basic covariates are included in the target logistic regression model. The proposed fine-tuning approach achieves performance comparable to models trained directly on the target population, indicating its effectiveness in leveraging scientific knowledge for domain adaptation. Notably, when partial target data are available, fine-tuning with only 10\% of the target data achieves performance similarly to that obtained using the full source dataset.

In summary, the main contributions of this paper are:
\begin{enumerate}
    \item Extending the GRU-D framework to incorporate static features through static feature embeddings with attention weight.
    \item Demonstrating the validity of transfer prediction across heterogeneous populations using soft labels derived from summary-level scientific knowledge.
    \item Establishing benchmarks for fine-tuning performance when only partial target data are available.
\end{enumerate}

These findings suggest the potential that even limited summary-level information or external scientific knowledge can enhance model predictive performance, thereby reducing the reliance on large-scale data collection in underrepresented populations. These results also support further evaluation of the proposed approach using real-world external summary sources.

\newpage
\bibliography{bib_file}

\section*{Acknowledgements}

The authors would like to thank the members and study participants of the Pregnancy Risk, Infant Surveillance, and Measurement Alliance (PRISMA) for their time and efforts. This study would not be possible without the support of the Bill \& Melinda Gates Foundation, specifically from Laura Lamberti and Richard Zong.  The conclusions and opinions expressed in this work are those of the author(s) alone and shall not be attributed to the Foundation. Under the grant conditions of the Foundation, a Creative Commons Attribution 4.0 License has already been assigned to the Author Accepted Manuscript version that might arise from this submission.

\section*{Author Contributions Statement}
Y.W., Q.P. and F.J. developed the transfer learning approach; E.R.S., Z. H. and Q. P. wrote the global health background and clinical interpretations; Y.W. carried out the simulation and real data analyses; all authors contributed to the data collection and writing of the manuscript.

\section*{Funding Information}
This work is funded by the Bill and Melinda Gates Foundation INV-041999 (PI: Emily R. Smith).
\section*{Data Availability}
The data that support the findings of this study are collected by the PRISMA consortium, but restrictions apply to the availability of these data, which were used under license for the current study and so are not publicly available. The data are, however, available upon request and with the permission of the PRISMA consortium.
\section*{Human Subjects Research}
All methods were carried out in accordance with relevant guidelines and regulations. The PRISMA MNH study was approved by the George Washington University’s Committee on Human Research (IRB: FWA00005945) on September 30, 2022 and received local and national ethical approval in Pakistan (Aga Khan University ERC 2022-5920-22763, and Pakistan National Bioethics Committee 4-87/NBC-58/8/22/337), Kenya (KEMRI Scientific and Ethics Review Unit KEMRI/SERU/CGHR/04/10/358/4166; Liverpool 23-020), Zambia (University of Zambia Biomedical Research Ethics Committee: 016-04-14 and University of North Carolina Chapel Hill Office of Human Research Ethics: 356795), Ghana (Kintampo Health Research Centre Institutional Ethics Committee (IEC) FWA00011103; Ref 0004854 and Ghana Health Service Ethics Review Committee FWA00020025), Vellore, India (Christian Medical College Vellore Office of Research IRB No 14553), and Hodal, India (Ethics Review Committee, Society for Applied Studies, SAS/ERC/ReMAPP Study/2022). Informed consent was obtained from participants at the time of original data collection. Work in this manuscript is covered by the IRB and informed consents of the overall PRISMA MNH study.

\end{document}